\documentclass{moriond}
\usepackage{amsmath,amsthm,amsfonts,amssymb,amscd,braket}

\def\be{\begin{equation}}
\def\ee{\end{equation}}
\def\bea{\begin{eqnarray}}
\def\eea{\end{eqnarray}}

\newcommand{\SO}[1]{\ensuremath{\mathrm{SO}(#1)}}
\newcommand{\SU}[1]{\ensuremath{\mathrm{SU}(#1)}}
\newcommand{\U}[1]{\ensuremath{\mathrm{U}(#1)}}

\newcommand{\mfrac}{\genfrac{}{}{0.34pt}1}
\newcommand*{\rep}[2][]{\ensuremath{{\boldsymbol{#2}#1}}} 

\newcommand{\Photo}{}

\begin{document}
\vspace*{4cm}
\renewcommand*{\thefootnote}{\fnsymbol{footnote}}
\title{Electroweak hierarchy from conformal and custodial symmetry \\[0.1cm] 
``Custodial Naturalness''
\footnote[1]{%
Contribution to the 2025 Electroweak session of the 59th Rencontres de Moriond.
}
}

\author{
\renewcommand*{\thefootnote}{\alph{footnote}}
Andreas Trautner\,\footnote{
\href{mailto:trautner@cftp.ist.utl.pt}{trautner@cftp.ist.utl.pt}
}
}

\address{CFTP, Departamento de F\'isica, Instituto Superior T\'ecnico, Universidade de Lisboa \\ Avenida Rovisco Pais 1, 1049 Lisboa, Portugal}

\maketitle\abstracts{
I introduce the idea of ``Custodial Naturalness'' to explain the origin of the electroweak~(EW) scale hierarchy. Custodial Naturalness is based on classical scale invariance as well as an extension of the Standard Model~(SM) scalar sector custodial symmetry. In a minimal realization, this requires a single new complex scalar field charged under a new U(1) gauge symmetry, which partially overlaps with $B-L$. Classical scale invariance and the high-scale scalar sector SO(6) custodial symmetry are radiatively broken by quantum effects that generate a new intermediate scale by dimensional transmutation. The little hierarchy problem is solved because the Higgs boson arises as pseudo-Nambu-Goldstone boson~(pNGB) of the spontaneously broken SO(6) custodial symmetry. The minimal realization of Custodial Naturalness has the same number of parameters as the SM and predicts testable new physics in the form of a heavy $Z'$ as well as a light but close-to invisible dilaton.}

\renewcommand*{\thefootnote}{\alph{footnote}}
\section{Hierarchical Scales in QFT and in the Standard Model}
The SM itself does not have a hierarchy problem because it is a one scale theory~\cite{Bardeen:1995kv}. If there are other hard scales next to the EW scale, the physical Higgs mass should be sensitive to these scales, giving rise to the EW hierarchy problem. It is not entirely clear whether this problem persists if all scales are generated dynamically. At least, we are not aware of any theorem which proves this conclusively (for concrete arguments, however, see~\cite{MarquesTavares:2013szc}). Well-accepted scenarios in which the Higgs mass sensitivity to high scales is lifted are supersymmetry or composite Higgs models. However, neither is Nature close to supersymmetric, nor do LHC measurements hint at deviations from an elementary Higgs boson. Also, no top partners have been observed which are a common requirement to both of these protection mechanisms. On the other hand, the SM \textit{is} close to being classically scale invariant with the only breaking term being the Higgs field quadratic~$\mu_{\mathrm{EW}}^2|H|^2$. Hence, we think it is worth revisiting the question if the only explicit mass scale in the SM could be generated dynamically, e.g.~via a quantum critical scale generation, i.e.~``dimensional transmutation'' \'a la Coleman-Weinberg~(CW)~\cite{Coleman:1973jx}. In the SM alone, this parametrically requires $m_t\lesssim m_Z$ and $m_h\lesssim 10\,\mathrm{GeV}$~\cite{Weinberg:1976pe,Gildener:1976ih}, which has long been excluded experimentally. Nonetheless, experimental data within $1\sigma$ points to a quantum critical value of the Higgs self-coupling at a high scale~\cite{CMS:2019esx,Hiller:2024zjp} (see talk by Daniel Litim), suggesting that dynamical scale generation may actually not be a bad idea. An experimentally not excluded way to realize this is to introduce an extended scalar sector that undergoes dimensional transmutation generating a scale $\Lambda_{\mathrm{CW}}$ which dynamically induces the EW scale via a Higgs portal coupling $\lambda_p|H|^2|\Phi|^2$~\cite{Hempfling:1996ht,Meissner:2006zh}. 
However, this typically introduces a little hierarchy problem. The mechanism of ``Custodial Naturalness'' improves on this situation and provides us with a scenario that features $(i)$ a technically natural suppression of EW scale, $(ii)$ only elementary (non-compositness) fields in a perturbative regime, and $(iii)$ a marginal top Yukawa coupling as in the SM with no requirement of top partners.

\section{General Idea of Custodial Naturalness}
The following assumptions underlie the mechanism of Custodial Naturalness: 1.\,Classical scale invariance; 2.\,Existence of new scalar degrees of freedom, here a complex field $\Phi$; 3.\,A mechanism to trigger the quantum critical scale generation for $\Phi$, here a new $\U{1}_{\mathrm{X}}$ gauge symmetry; 4.\,High-scale custodial symmetry (CS) of the scalar potential, here \SO{6}. 

$H$ and $\Phi$ then form a $\rep{6}$-plet of \SO{6}, with a custodially symmetric potential
\be
  \qquad\qquad \Big. V(H,\Phi)~=~\lambda\left(|H|^2+|\Phi|^2\right)^2\;\qquad\quad\text{at}~~\mu=\Lambda_{\mathrm{high}}= M_{\mathrm{Pl}}\,.
 \ee
Both, scale invariance and custodial symmetry are explicitly broken and this modifies the high-scale tree-level potential by quantum effects. Dimensional transmutation gives rise to a vacuum expectation value (VEV) of the $H$-$\Phi$ system and spontaneous breaking of \SO{6}. In addition, \SO{6} is explicitly broken by the top Yukawa coupling $y_t$, hypercharge gauge coupling $g_{\mathrm{Y}}$, new $\U{1}_{\mathrm{X}}$ gauge coupling $g_{\mathrm{X}}$, $\U{1}_{\mathrm{X}}$-$\U{1}_{\mathrm{Y}}$ gauge-kinetic mixing~\footnote{
$g_{12}:=\epsilon g_Y/\sqrt{1-\epsilon^2}$ with field strength mixing $\epsilon\,F^{\mu\nu}F'_{\mu\nu}$.}
$g_{12}$, and potentially new Yukawa couplings of $\Phi$. The six real scalar degrees of freedom amount to four would-be NGBs (eaten by the SM EW bosons and new $Z'$), one massive dilaton (pNGB of spontaneously broken and anomalous scale invariance) and one massive pNGB of spontaneously broken custodial symmetry that very closely resembles the SM Higgs boson. 

\begin{figure}
\centerline{\includegraphics[width=0.65\linewidth]{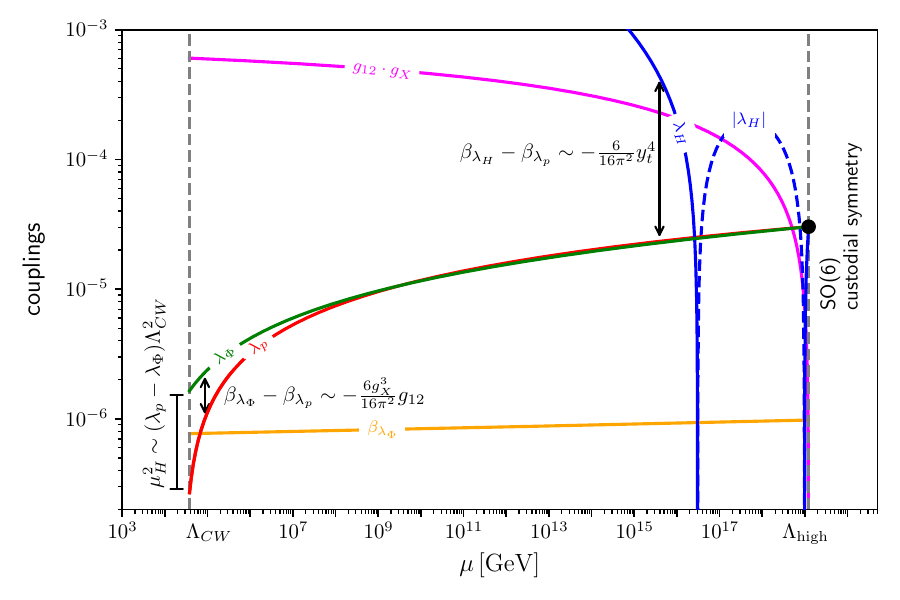}}
\caption[]{
Renormalization group running of parameters for an actual benchmark point (BP, labeled by black star on other plots). We indicate the approximate splitting of beta functions along the flow.}
\label{fig:running}
\end{figure}
Below $\Lambda_{\mathrm{high}}$ we use the modified tree-level potential
\begin{equation}
\quad V_{\text{tree}}(H,\Phi)~=~\lambda_H |H|^4 +2\,\lambda_p|\Phi|^2 |H|^2+ \lambda_\Phi|\Phi|^4\;,
\end{equation}
with couplings that run as illustrated by our benchmark point in Fig.~\ref{fig:running}. The dominant source of CS breaking is $y_t$, which drives $\lambda_H\gg\lambda_{p,\Phi}$ to its SM value. This automatically leads to  
$\langle H\rangle\ll \langle\Phi\rangle$. 
The masses of physical bosons are to first approximation given by
\begin{align}
&\text{$Z'$:}& m^2_{Z'}&~\approx~g_X^2v_{\Phi}^2\;,\\
&\text{Dilaton:}& m^2_{h_\Phi}&~\approx~\beta_{\lambda_\Phi}\,v^2_{\Phi}~\approx~\frac{3\,g_X^4}{8\pi^2}\,v^2_{\Phi}\;, \\
&\text{pNGB Higgs:}& m^2_h~&~\approx~2\left[\lambda_\Phi \left(1+\frac{g_{12}}{2\,g_X}\right)^2-\lambda_p\right]\,{v^2_{\Phi}}\;,
\end{align}
where $\langle\Phi\rangle=\frac{v_\Phi}{\sqrt{2}}$, $\langle H\rangle=\frac{v_H}{\sqrt{2}}$, and up to small mixing effects $h_\Phi\subset\Phi$ and $h\subset H$.

The close proximity and stability of $\lambda_{p}\sim\lambda_{\Phi}$ is explained by the remaining custodial symmetry. Crucially, $\mu_{\mathrm{EW}}^2\sim \left[\lambda_p-\lambda_\Phi\right]\,v_{\Phi}^2$, implying that the Higgs mass and EW scale are \textit{custodially} suppressed relative to the intermediate scale~$v_\Phi$ of spontaneous scale and custodial symmetry violation. The SM relation $v_H^2\approx m^2_h/2\lambda_H$ is maintained. In this class of models one does not have to work to make the Higgs light, but one has to work to make it heavy enough: Getting the EW scale high enough typically \textit{requires} an additional BSM source of CS violation.  In the simplest case this is $g_X$ and the gauge kinetic mixing $g_{12}$.

\section{A Minimal Model}
\begin{table}[t]
\caption[]{Field content of the Custodial Naturalness model discussed here. We display the $\U{1}_{\mathrm{X}}$ charges resulting for our benchmark choice $q^{\mathrm{B-L}}_\Phi=-\mfrac{1}{3}$.}
\label{tab:model}
\vspace{0.4cm}
\begin{center}
\renewcommand{\arraystretch}{1.35}
\scalebox{0.9}{
	\begin{tabular}{|ccccc|}
	\hline
		Field & \#Gens.  & $\SU{3}_{\mathrm{c}}\!\times\!\SU{2}_{\mathrm{L}}\!\times\!\U{1}_{\mathrm{Y}}$\!&\! $\U{1}_{\mathrm{X}}$\! & ~~~~$\U{1}_{\mathrm{B-L}}$ \\ 
 		\hline 
 		$Q$ & $3$  & $(\rep{3},\rep{2},+\mfrac{1}{6})$  & $-\frac{2}{3}$ & ~~~~$+\frac{1}{3}$ \\
		$u_R$ & $3$  & $(\rep{3},\rep{1},+\mfrac{2}{3})$  & $+\mfrac{1}{3}$ & ~~~~$+\mfrac{1}{3}$ \\
		$d_R$ & $3$ & $(\rep{3},\rep{1},-\mfrac{1}{3})$ & $-\mfrac{5}{3}$ & ~~~~$+\mfrac{1}{3}$ \\
		$L$ & $3$ & $(\rep{1},\rep{2},-\mfrac12)$ & $+2$ & ~~~~$-1$ \\
		$e_R$ & $3$ & $(\rep{1},\rep{1},-1)$  & $+1$ & ~~~~$-1$ \\
		$\nu_R$ & $3$ & $(\rep{1},\rep{1},\phantom{-}0)$ & $+3$ & ~~~~$-1$ \\
        \hline
		$H$ & $1$ & $(\rep{1},\rep{2},+\mfrac12)$ & $+1$ & ~~~~$\phantom{-}0$ \\
		$\Phi$ & $1$ & $(\rep{1},\rep{1},\phantom{-}0)$ & $+1$ & ~~~~$q^{\mathrm{B-L}}_\Phi=-\mfrac{1}{3}$ \\[1pt]
		\hline
	\end{tabular}
}
\end{center}
\end{table}\enlargethispage{0.5cm}
The field content of a minimal extension that realizes the idea of Custodial Naturalness is shown in Tab.~\ref{tab:model} (incl.~SM fields+3 gens.\ of $\nu_R$, excl.\ gauge bosons). There are two new physical fields: the dilaton $h_{\Phi}$ and $Z'$ gauge boson. The only free parameter of the charge assignment is $q^{\mathrm{B-L}}_{\Phi}$, the $B-L$ charge of $\Phi$. The direction of the gauged $\U{1}_{\mathrm{X}}$ as a liner combination of hypercharge and $B-L$ is defined such that $H$ and $\Phi$ have the same $\U{1}_{\mathrm{X}}$ charge. Hence, 
\be
Q^{(\mathrm{X})}~\equiv~2\,Q^{(\mathrm{Y})}+\frac{1}{q^{\mathrm{B-L}}_\Phi}\,Q^{(\mathrm{B-L})}\;.
\ee\enlargethispage{0.5cm}
Roughly, we have the constraints $\mfrac13 \lesssim |q^{\mathrm{B-L}}_\Phi|\lesssim\mfrac{5}{11}$ to allow for sufficient amount of CS violation (low end) but avoid a Landau pole below $M_{\mathrm{Pl}}$ (high end).\footnote{ 
This excludes a value of $q^{\mathrm{B-L}}_\Phi=-2$ for our purposes, which is commonly chosen in the ``classical conformal extension of minimal $B-L$ model''~\cite{Iso:2009ss,Iso:2009nw} (that happens to be very similar to our model). Interestingly, the range contains the special value $q^{\mathrm{B-L}}_\Phi=-\mfrac{16}{41}$ that leads to absence of gauge-kinetic mixing even at one loop.} 
To display the charges and discuss experimental constraints we fix $q^{\mathrm{B-L}}_\Phi=-\mfrac{1}{3}$ as a benchmark value. 

Assuming the absence of gauge-kinetic mixing at the high scale (which is motivated by enhancing CS), this model has the same number of parameters as the SM ($G_{\mathrm{F}}$ and $m_h$ get traded for $\lambda$ and $g_X$). Turning on $g_{12}$ at the high scale serves as a proxy for us to investigate the opening-up of viable parameter space in presence of additional sources of CS breaking.

\section{Experimental Tests}
Fixing the known SM phenomenology, most importantly EW scale, Higgs and top masses, there are no free parameters 
and the properties (masses, mixings, decay rates) of the new particles $Z'$, $h_\Phi$ are predictions of the model. Hence, prime targets to test this model are direct $Z'$ searches at current and future energy-frontier colliders as well as dilaton production at future Higgs factories.

\begin{figure}
\begin{minipage}{0.5\linewidth}
\centerline{\includegraphics[width=1.0\linewidth]{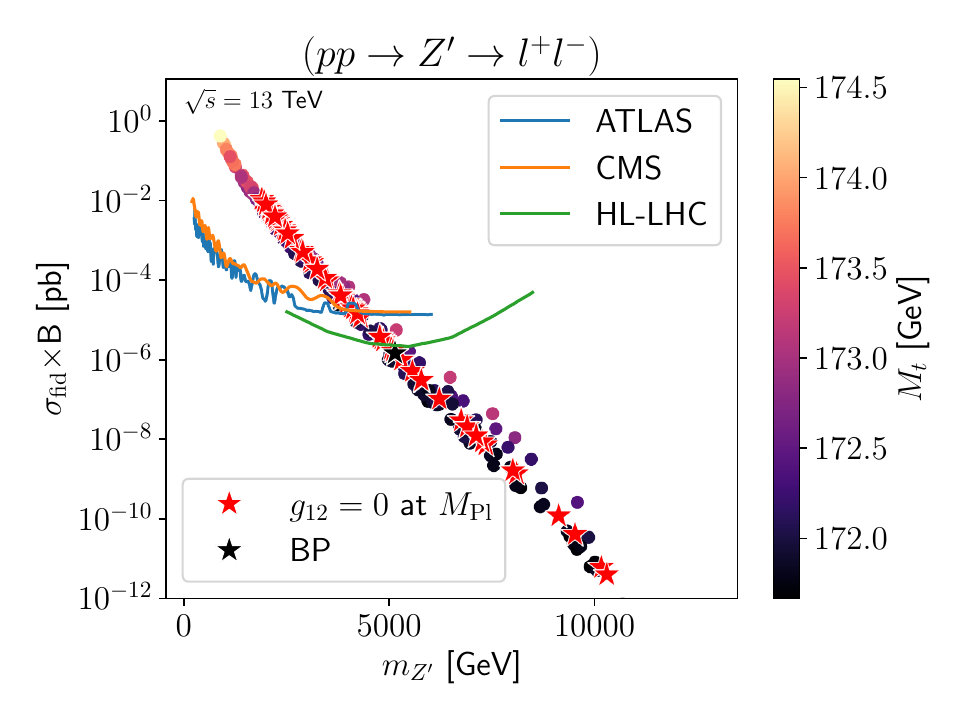}}
\end{minipage}\hfill
\begin{minipage}{0.5\linewidth}
\centerline{\includegraphics[width=1.0\linewidth]{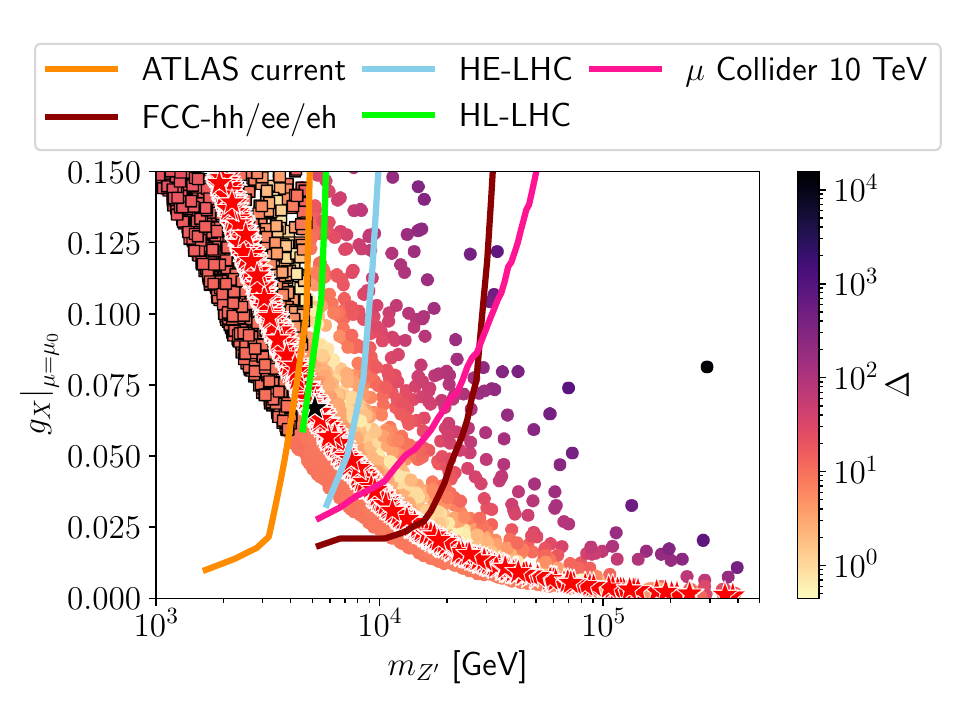}}
\end{minipage}
\caption[]{
Left: Fiducial cross section times branching ratio for $Z'$ at the LHC and experimental constraints from ATLAS~\cite{ATLAS:2019erb} and CMS~\cite{CMS:2021ctt} as well as projections of HL-LHC constraints. Right: Model parameter space and projected future limits adopted from Ref.~\cite{EuropeanStrategyforParticlePhysicsPreparatoryGroup:2019qin} for the case of a hypercharge universal $Z'$.}
\label{fig:limits}
\end{figure}
A substantial region of otherwise viable parameter space is already excluded by ATLAS and CMS $Z'\rightarrow l^+l^-$ dilepton resonance searches~\cite{ATLAS:2019erb,CMS:2021ctt}, see Fig.~\ref{fig:limits}~(left), which gives a constraint $m_{Z'}\gtrsim 4\,\mathrm{TeV}$ (dijet resonance constraints are only a little bit weaker). Effects on EW precision at zeroth order correspond to a shift in the $Z$ mass, $\Delta m_Z\propto-m_Z \langle H\rangle^2/(2\langle\Phi\rangle^2)$, demanding $\langle\Phi\rangle\gtrsim 18\,\mathrm{TeV}$ which is less restrictive than than direct constraints. A full EW precision fit in a more refined future analysis could reveal possible BSM correlations in EW precision observables. Dilaton-Higgs mixing is typically present at the level of $\sin^2\theta\sim10^{-5}$, but is currently not constrained beyond $\sin^2\theta\lesssim10^{-2}$ at best.
The dilaton phenomenology essentially corresponds to that of a singlet scalar SM extension, where operators containing the new field are given by $\mathcal{O}_{h_\Phi}\approx \sin\theta \times \mathcal{O}^{\mathrm{SM}}_{h\rightarrow h_\Phi}$. This implies that dilaton branching ratios are the same as for a comparable-mass SM Higgs, with absolute decay rates rescaled by $\sin^2\theta$. Deviations from this simplified scenario arise because the dilaton additionally couples to the trace anomaly. These effects are suppressed by factors $v_H/v_\Phi$. Hence, they only matter for a precision measurement of dilaton branching ratios where, however, they distinguish the dilaton from the singlet extension. There is an interesting window in the parameter space where the dilaton-Higgs mixing is large enough to produce a number of dilatons at future Higgs factories, but small enough such that the dilaton is long-lived enough to give a background-free displaced vertex signature. 

\begin{figure}
\begin{minipage}{0.5\linewidth}
\centerline{\includegraphics[width=1.0\linewidth]{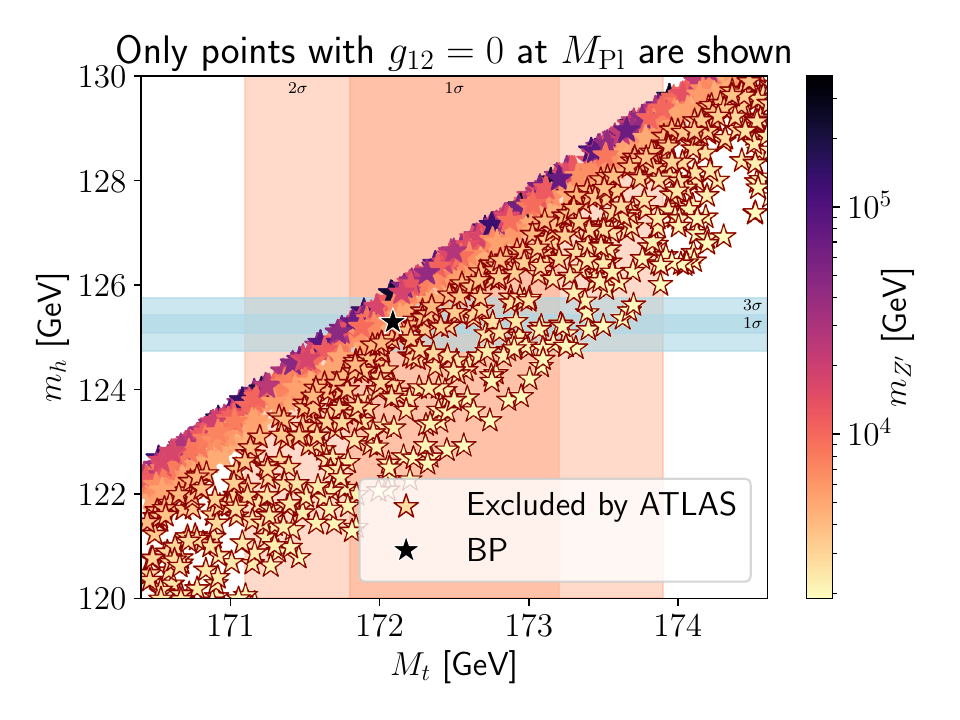}}
\end{minipage}\hfill
\begin{minipage}{0.5\linewidth}
\centerline{\includegraphics[width=1.0\linewidth]{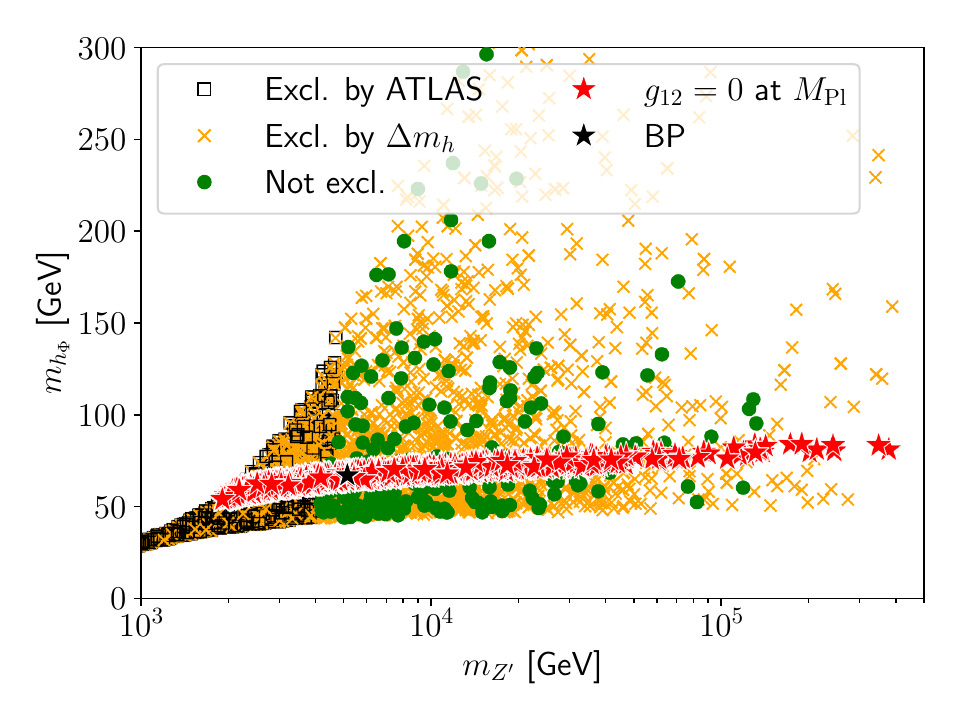}}
\end{minipage}
\caption[]{Correlations of top pole mass and Higgs mass~(left), as well a between $m_{Z'}$ and dilaton $m_{h_{\Phi}}$~(right).}
\label{fig:correlations}
\end{figure}
For details of the parameter scan we refer to~\cite{deBoer:2024jne,deBoer:2025oyx}. In Fig.~\ref{fig:correlations}~(left) we show the correlation between top pole mass $M_t$ and Higgs mass for all points that reproduce a correct EW scale in case $g_{12}|_{M_{\mathrm{Pl}}}=0$. Viable values for $M_t$ and $m_h$ can only be reproduced in this model if the top weights in at the lower end of its currently allowed $1\sigma$ mass range. Fig.~\ref{fig:correlations}~(right) shows the correlation between BSM particle masses $m_{Z'}$ and $m_{h_\Phi}$. Here we also show points with $g_{12}|_{M_{\mathrm{Pl}}}\neq0$ to demonstrate the opening-up of the parameter space in the presence of additional sources of CS violation. In Fig.~\ref{fig:limits}~(right) we show projections for the sensitivity of future colliders. In color we show fine tuning (or absence thereof) in $\langle H\rangle/\langle \Phi\rangle$ for parameter points that satisfy all constraints.
Here we use a Barbieri-Giudice~\cite{Barbieri:1987fn} measure $\Delta:=\max_{g_i}\,\left|
 \partial\,\ln(\langle H\rangle/\langle \Phi\rangle)/\partial\ln g_i\right|$. This automatically subtracts the shared common sensitivity of VEVs to high-scale variations of $g_i$ which is meaningless in scenarios of dimensional transmutation~\cite{Anderson:1994dz}, see~\cite{deBoer:2024jne,deBoer:2025oyx} for details. 

\enlargethispage{1cm}
\section{Other Signatures and Model Variations}
The CW phase transition is generically of first order, hence, gives rise to gravitational wave~(GW) signals. 
In fact, the ``minimal conformal $B-L$ model''~\cite{Iso:2009ss,Iso:2009nw} is a prototype model for a transition with strong supercooling and abundant gravitational wave production from bubble collisions, see e.g.~\cite{Ellis:2020nnr} and the talk by Kamila Kowalska. However, quantitative predictions for GW signal of this model have yet to be worked out. 

Custodial Naturalness is reasonably stable under variation of boundary conditions, charge assignments, and/or addition of extra fields~\cite{deBoer:2024jne,deBoer:2025oyx}. Many variations and extensions are possible, as already known from the conformal $B-L$ model~\cite{Iso:2009ss,Iso:2009nw} (see also references in~\cite{deBoer:2024jne}). Additional fermions can provide ingredients for neutrino mass generation, be part of the dark matter, or can be used to ``cure'' the SM vacuum instability~\cite{Oda:2015gna,Das:2015nwk}.\footnote{%
We emphasize that experiment actually shows that Nature sits on the meta-stable critical line at $1\sigma$~\cite{CMS:2019esx,Hiller:2024zjp}. We think that the vacuum meta-stability and associated quantum criticality may not be a bug but a feature.}

There is a variant of the scenario of Custodial Naturalness that is more minimal than the extensions discussed here in terms of field content, but contains at least one more parameter. 
This is the scenario of minimal ``Scalar Custodial Naturalness''~\cite{deBoer:2025xx}. It extends the SM$+3\nu_R$ by two real scalar fields $\phi$ and $S$ and the assumption of $\SO{5}$ custodial boundary conditions in $\rep{5}=H+\phi$. This can lead to CW quantum critical scale generation without a new gauge sector, while having an automatic scalar DM candidate in $S$~\cite{deBoer:2025xx}. Unfortunately, due to the absence of a $Z'$ signature, this model is much harder to test than the \SO{6} scenario discussed here.

There is an additional twist possible in all scenarios discussed here: For a limited range of CS boundary condition scales $\Lambda_{\mathrm{high}}$, the SM itself has sufficient radiative CS violation to viably generate the EW scale from splitting $\lambda_p-\lambda_\Phi$ without the need of a BSM source of CS breaking. In this case, the scale of CS can be predicted and one typically finds~\cite{deBoer:2025xx} $\Lambda_{\mathrm{high}}\approx 10^{11}\,\mathrm{GeV}$.

\section{Conclusions}
To summarize, quantum criticality may play an important role in generation of the EW scale. Classical scale invariance and an extended custodial symmetry provide a new mechanism to explain a large scale separation and also resolve the little hierarchy problem without fine tuning. The minimal model discussed here has the same number of parameters as the SM. It predicts a light scalar dilaton
with a mass around $m_\Phi\sim75\,\mathrm{GeV}$, as well as a new heavy $Z'$ with a mass in the range of $4-100\,\mathrm{TeV}$.
Probably the most accessible signature of our scenario is the expectation that the top pole mass should fall into the lower end of its currently allowed $1\sigma$ interval.
The mechanism of Custodial Naturalness and the associated predictions are reasonably stable under extensions of the minimal scenario, e.g.\ to accommodate neutrino masses or particle DM.
This is a well motivated model to search for at future colliders, Higgs factories, and gravitational wave observatories. Our discussion also provides a good starting point for further extensions, for example, to minimize the number of fields as in
minimal scalar Custodial Naturalness, or for exploring flavor dependent fermion charge assignments.

\section*{Acknowledgments}
I am grateful to my collaborators Thede de Boer and Manfred Lindner, as well as to the participants of the Moriond meeting for stimulating discussions of the ideas presented in this talk. This work is supported in part by the Portuguese Funda\c{c}\~ao para a Ci\^encia e a Tecnologia (FCT) through project \href{https://doi.org/10.54499/2023.06787.CEECIND/CP2830/CT0005}{2023.06787.CEECIND} and contract \href{https://doi.org/10.54499/2024.01362.CERN}{2024.01362.CERN}, partially funded through POCTI (FEDER), COMPETE, QREN, PRR, and the EU.

\section*{References}
\bibliographystyle{moriond}
\bibliography{bib}

\end{document}